\documentclass[final,12pt,5p,twocolumn]{elsarticle}

\usepackage{amssymb}
\usepackage{amsmath}

\usepackage[final]{changes}
\usepackage{ulem}
\usepackage{xcolor}

\usepackage{tabularx,ragged2e}
\newcommand{\HY}{\hyphenpenalty=25\exhyphenpenalty=25} 
\newcolumntype{Z}{>{\HY\RaggedRight\arraybackslash\hspace{0pt}}X} 

\usepackage[hyphens]{url}
\usepackage{hyperref} 
\hypersetup{breaklinks=true}
\usepackage{caption}
\usepackage{booktabs}
\usepackage{multirow}
\usepackage{tabularx}	
\usepackage{color,colortbl}
\usepackage{subcaption}
\usepackage{graphicx}   

\makeatletter
\newcommand{\ion}[2]{%
  #1$\;$%
  \if b\expandafter\@car\f@series\relax\@nil
    \begingroup 
      \sbox0{\rmfamily\mdseries\textsc{v}}%
      \resizebox{!}{\ht0}{\rmfamily\@Roman{#2}}%
    \endgroup
  \else
    \textsc{\rmfamily\@roman{#2}}%
  \fi
}
\makeatother

\makeatletter
\def\@author#1{\g@addto@macro\elsauthors{\normalsize%
    \def\baselinestretch{1}%
    \upshape\authorsep#1\unskip\textsuperscript{%
      \ifx\@fnmark\@empty\else\unskip\sep\@fnmark\let\sep=,\fi
      \ifx\@corref\@empty\else\unskip\sep\@corref\let\sep=,\fi
      }%
    \def\authorsep{\unskip,\space}%
    \global\let\@fnmark\@empty
    \global\let\@corref\@empty  
    \global\let\sep\@empty}%
    \@eadauthor={#1}
}
\makeatother

\journal{Journal Name}

\begin{document}
\begin{frontmatter}



\title{Detecting Rug Pulls in Decentralized Exchanges: Machine Learning Evidence from the TON Blockchain}


\author[MIPT]{Dmitry Yaremus}
\ead{iaremus.dk@phystech.edu} 

\author[HSE]{Jianghai Li}
\ead{t-li@edu.hse.ru}

\author[MIPT]{Alisa Kalacheva}
\ead{alisa.kalacheva@skoltech.ru}


\author[IR]{Igor Vodolazov}
\ead{allcryptodata@proton.me}

\author[Sk,HSE]{Yury Yanovich}
\ead{{Corresponding author*}{y.yanovich@skoltech.ru}}

\affiliation[MIPT]{organization={Moscow Institute of Physics and Technology},
            city={Moscow 141701},
            country={Russia}}

\affiliation[HSE]{organization={Faculty of Computer Science, HSE University},
            city={Moscow 109028},
            country={Russia}}

\affiliation[MSU]{organization={Moscow State University},
            city={Moscow 119991},
            country={Russia}}

\affiliation[Sk]{organization={Skolkovo Institute of Science and Technology},
            city={Moscow 121205},
            country={Russia}}

\affiliation[IR]{organization={Independent Researcher},
            city={Barcelona 08001},
            country={Spain}}

\begin{abstract}
This paper presents a machine learning framework for the early detection of rug pull scams on decentralized exchanges (DEXs) within The Open Network (TON) blockchain. TON's unique architecture, characterized by asynchronous execution and a massive web2 user base from Telegram, presents a novel and critical environment for fraud analysis. We conduct a comprehensive study on the two largest TON DEXs, Ston.Fi and DeDust, fusing data from both platforms to train our models. A key contribution is the implementation and comparative analysis of two distinct rug pull definitions--TVL-based (a catastrophic liquidity withdrawal) and idle-based (a sudden cessation of all trading activity)--within a single, unified study. We demonstrate that Gradient Boosting models can effectively identify rug pulls within the first five minutes of trading, with the TVL-based method achieving superior AUC (up to 0.891) while the idle-based method excels at recall. Our analysis reveals that while feature sets are consistent across exchanges, their underlying distributions differ significantly, challenging straightforward data fusion and highlighting the need for robust, platform-aware models. This work provides a crucial early-warning mechanism for investors and enhances the security infrastructure of the rapidly growing TON DeFi ecosystem.
\end{abstract}

\begin{keyword}
Scam Detection \sep Data Fusion \sep Machine Learning \sep Blockchain \sep DEXs \sep TON 


\end{keyword}

\end{frontmatter}

\section{Introduction}

The Open Network \cite{TON2018} was originally conceived and developed by Telegram, and is now independently operated by the TON Foundation. It is a high-performance decentralized platform designed to support large-scale decentralized applications (DApps) \cite{Zheng2023} and smart contracts \cite{Zheng2020}. The number of users in the TON ecosystem has increased dramatically, with monthly active users increasing to 4.64 million. However, some users are unaware of blockchain risks and are easy targets for fraud and hacker attacks \cite{blockchainpoweredds2024}. This research focuses on smart contracts related to decentralized exchanges (DEXs) platforms \cite{Mohan2022}, such as Ston.Fi \cite{stonfi-api} and DeDust \cite{dedust}. In DEXs, tokens quickly lose value due to vulnerabilities in smart contract code, defects in liquidity mechanisms \cite{kosmosis-rug-pull, sok-rug-pull, prevalence-rugpulls}, lack of regulation, and anonymity. A common scam encountered by users is rug pull, in which the project party suddenly withdraws funds and runs away after attracting sufficient liquidity. Mantra (OM) is considered the biggest rug pull of the year, with losses amounting to \$5.52 billion \cite{om2025}. This study applies machine learning to DEXs in TON for early rug pull detection, which is capable of achieving efficient early warnings solely based on minute-level data from the initial stages of token transactions.

The high failure rate of the cryptocurrency market shows that the early window period is the stage of highest risk \cite{jin2022dualchannelearlywarningframework}. The early window detection of tokens is conducive to protecting user assets and provides a means of prevention \cite{Cheng2023}. Although the machine learning method cannot capture the new type of rug pull with 100\% accuracy, it has played an important reference role for traders in the early stages. This study aims to use machine learning to discover and detect the early behavior patterns of rug pull on different DEXs data. It will also develop the early window size of the rug pull token, and analyze the data of all available tokens to find appropriate strategies. The results are helpful in understanding different early window strategies and collecting data from different sources to improve the machine learning model.

This study makes the following key contributions to the field of on-chain security and rug pull detection:
\begin{enumerate}
    \item \textbf{Rug Pull Analysis for the TON Blockchain Ecosystem:} We present an extensive analysis of rug pull scams on the TON blockchain, addressing its unique ecosystem defined by asynchronous transaction execution and a large, less crypto-native audience migrating from Web2 via Telegram. Our work provides a foundational dataset and benchmark for security research in this emerging environment.
    \item \textbf{Unified Evaluation of TVL and Idle-Based Rug Pull Definitions:} We implement, validate, and directly compare the two primary rug pull detection methodologies from the literature -- TVL-based and idle-based -- within a single framework. This comparative analysis provides a comprehensive view of scam dynamics, showing that the TVL method achieves superior AUC while the idle method is optimal for maximizing recall.
    \item \textbf{Analysis of Cross-DEX Data Fusion Viability:} We investigate the viability of data fusion techniques for combining data from two major TON DEXs (Ston.Fi and DeDust). A critical finding is that while a consistent feature set can be engineered across platforms, their underlying statistical distributions differ significantly. This insight is vital for future multi-platform studies, indicating that models must account for such domain shifts.
\end{enumerate}

The rest of the paper is organized as follows. Section \ref{section:Related Work} provides an overview of the related work. Section \ref{section:Research Workflow} formalizes the methodology, defines target variables using the Idle and TVL approaches in \ref{subsection:Target variable definition (Idle vs. TVL)},  establishes evaluation criteria in \ref{subsection:Criteria for Model Evaluation}, and shows details of cross-platform data fusion techniques in \ref{subsection:Data Fusion}. Section \ref{section:Data Mining} covers the data mining in this study, including the collection of data from Ston.Fi and DeDust in \ref{subsection:Data Collection}, discusses the characterization of the data set in \ref{subsection:Dataset Description}, shows the details of feature engineering in \ref{subsection:Feature Engineering} , data labeling in \ref{subsection:Data Labeling} and preprocessing pipelines in \ref{subsection:Data_Preprocessing_Techniques}. Section \ref{section:Detection Model} detection model describes each sample preprocessing step in \ref{subsection:Data Preprocessing and Feature Engineering} and explains which machine learning model will be used and how to do tuning in \ref{subsection:Model and Hyperparameter Tuning}. Section \ref{section:Numerical_Results} presents performance benchmarks (AUC/Class 0 accuracy) and a comparative analysis of the efficacy of Idle and TVL. Section \ref{section:Discussions and Conclusions} provides the concluding remarks of the paper.

\section{Related Work}
\label{section:Related Work}

The evolution of blockchain and cryptocurrency has been accompanied by a constant struggle between innovation and new threats. In 2017-2018, there was an explosive growth in ICO (Initial Coin Offering), which led to the emergence of many scam projects and rug pull \cite{Howell2020}. In 2020-2024, DeFi and DEX came to the fore, as well as the creation of memecoins \cite{Stencel2023}, which increased trading volumes and attracted the attention of investors and scammers \cite{detecting-rug-pulls}.

Memcoins such as Notcoin, DOGS, DUREV on TON are becoming popular due to integration with messengers and social networks, which expands the audience, but also increases the risks for inexperienced users. Memcoins repeat the scenario of classic financial bubbles: rapid capitalization growth. In 2024, the memcoin market increased by 330\%, reaching \$140 billion.

Modern research offers comprehensive approaches to fraud detection that combine machine learning methods with blockchain features. Key areas include:

\begin{itemize}
\item \textbf{Ensemble classification methods}: The works \cite{Taher_Ameen_Ahmed_2024, rugpulls2024} demonstrate the effectiveness of a combination of algorithms (logistic regression, Isolation Forest) or XGBoost combined with SMOTE to detect suspicious transactions in Ethereum with an accuracy of up to 99\%.

\item \textbf{Graph analysis of transactions}: The study \cite{ijarsct2024phishing} proposes using node2vec to analyze Ethereum transactions, achieving an F1 score of 0.846 to detect phishing attacks.

\item \textbf{Real-time anomaly detection}: The framework combines clustering (k-means) and deep learning techniques to analyze blockchain streaming data \cite{Caron:2024zxk}.
\end{itemize}

A significant amount of research has been devoted to blockchain fraud, particularly schemes that are common in token trading. Several studies have looked at fraudulent schemes associated with ICO \cite{article_Liebau,Milind_Tiwari,chiu2022using}, for example, regression analysis showed that tokens that were listed on exchanges (previously considered a sign of a successful ICO) were more likely to be the target of fraud, in approximately 10.1\% \cite{hornuf2022initial}.

Exchanges and the tokens traded on them are also considered by researchers as a key source of vulnerability. Tokens with potential "backdoors" in their code are considered especially dangerous. Currently, various analysis tools are available that can be applied to centralized exchanges (CEX) and decentralized exchanges (DEX) \cite{ji2020, Durieux2020}.

The cryptocurrency market is vulnerable to external manipulation, which is one of the main principles of market vulnerability \cite{twomey2020fraud}. The authors highlight factors such as imperfect regulation, relative anonymity, low barriers to entry, and the lack of strict procedures to create exchanges. Exchange vulnerabilities account for a significant portion of cryptocurrency fraud cases. Between 2011 and 2017, 18 token exchanges were closed due to fraud \cite{xia2021trade}.

The rug pull scheme was studied by Bruno Mazorra \cite{mazorra2022rug}, who analyzed 28,000 tokens in Uniswap V2, of which 98\% were flagged as fraudulent in the data. His methodology is based on time series analysis using machine learning, estimating the price drop of a token.

An alternative method to detect token scams is the approach implemented in the TokenScout tool, which uses temporal graph neural networks to detect scams on the Ethereum blockchain \cite{wu2024tokenscout}. In another study, the authors used temporal characteristics and automated mechanisms to detect suspicious assets. The researchers developed scam recognition models adapted to work with different time intervals. Although large time windows can improve the accuracy of the models, it is essential to consider short periods to minimize delays in detecting rug pulls after they have occurred \cite{srifa2025rug}.

In the context of the analysis of rug pulls in TON, it is also worth mentioning the study \cite{cernera2023token}, which looked at token activity on the Ethereum and Binance Smart Chain (BSC) blockchains until 2022. According to the results, about 70\% of the addresses created only one token, while only 1\% of the addresses issued more than 18 tokens each. The token life cycle has become even shorter: almost half (49.7\%) of the assets disappear from the exchange within the first 4 hours after launch.

Memecoins, due to their high virality and low entry barrier, are attractive targets for attackers \cite{rugpulls2024}. This article focuses on the characteristics of memecoins, which have short life cycles and rapidly fluctuating activity. This research needs to explore machine learning methods for analyzing smart contract and transaction information and proposes the development of effective tools to protect users and strengthen the DeFi ecosystem's resilience to fraudulent schemes.

These findings confirm the relevance of this study. The more deeply the behavior of tokens on exchanges is studied and predicted, the more opportunities there are for safe and informed investments.

\section{Research Workflow}
\label{section:Research Workflow}
The goal of this research is to develop an automated solution for the early lifetime detection of fraudulent tokens (rug pulls) on decentralized blockchain exchanges (DEXs of TON) -- $Ston.Fi$ and $DeDust$. It protects investors' interests and enhances trust in the TON DeFi ecosystem by timely detecting rug pulls based on data from the first few minutes of trading. Specifically, it is necessary to predict whether a rug pull (fraudulent liquidity loss or cessation of activity) will occur within the next hour after the start of trading, based on data collected within 5 minutes of the start of DEX token trading. The solution should be implemented using machine learning methods and provide investors with recommendations on how to participate in new token trading.
The processing to detect rug pull on the TON blockchain includes the following steps:

\begin{itemize}
\item \textbf{Data Collection:} Obtaining information on transactions and liquidity pools for tokens on the $Ston.Fi$ and $DeDust$ DEXs through indexed data from the TON Foundation and the Dune.com service.
\item \textbf{Data Preprocessing:} Cleaning of gaps and zero values, scaling, and generating features.
\item \textbf{Feature Generation:} Formation of transaction, price, liquidity, time, and meta-features for each token.
\item \textbf{Data Labeling:} Defining the target variable (rug pull) using two approaches: Idle (no transactions for an hour) and TVL (a drop in TVL by more than 99\% from the maximum in the first hour).
\item \textbf{Training the models:} Applying machine learning algorithms to predict rug pulls.
\item \textbf{Validation and Evaluation:} Using cross-validation and test sets to evaluate the quality of the models.
\item \textbf{Analysis of Results:} Comparing the performance of models and approaches, analyzing the importance of features.
\end{itemize}

\subsection{Target Variable Definition: Idle vs TVL}
\label{subsection:Target variable definition (Idle vs. TVL)}

The study will use Idle and TVL approaches \cite{rugpulls2024, mazorra2022rug} to define rug pull, each corresponding to its target variable.

\subsubsection{Idle Approach} 
\label{subsubsection:Idle Approach}
\begin{figure}[!h]
    \centering
    \includegraphics[width=0.95\linewidth]{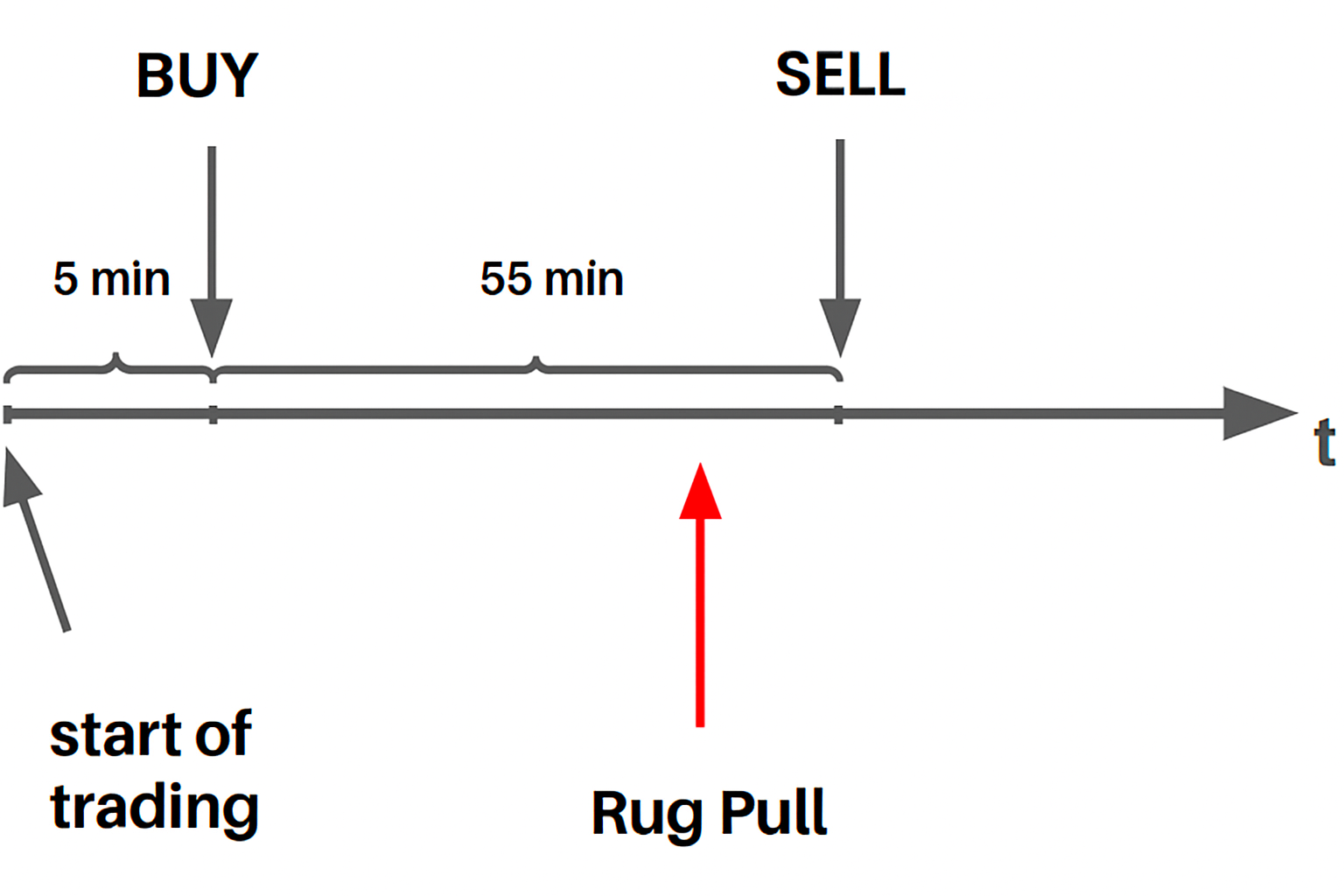}
    \caption{Visualization of the general statement of the problem}
    \label{fig:buy_sell}
\end{figure}

\begin{figure}[!h]
    \centering
    \includegraphics[width=0.95\linewidth]{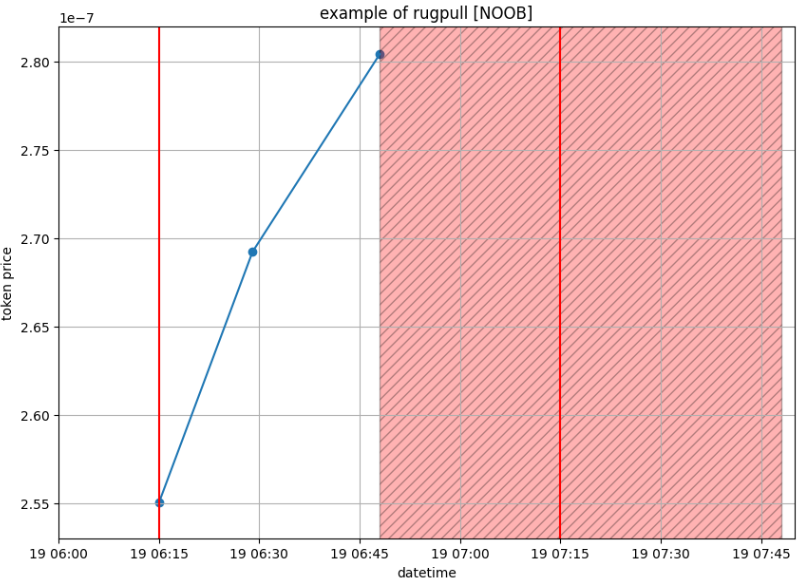}
    \caption{Example of Idle rug pull with \$NOOB token}
    \label{fig:noob}
\end{figure}

Rug pull is defined as a token that has no trades (buys/sells) within one hour of starting the trading. Figure \ref{fig:buy_sell} shows this method. It is recommended by decentralized exchange trading enthusiasts because it has good practical value, if there are no trades within an hour, the previously purchased tokens will not be able to be sold. Figure \ref{fig:noob} shows the \$NOOB token -- 3 trades within 30 minutes of the start of the trade, then the activity stops for a full hour. 

\subsubsection{TVL (Total Value Locked) Approach} 
\label{subsubsection:TVL (Total Value Locked) Approach}
A rug pull is defined as a drop in TVL of more than $p$\% from its peak value within the first hour of trading.

\begin{figure}[!h]
    \centering
    \includegraphics[width=0.95\linewidth]{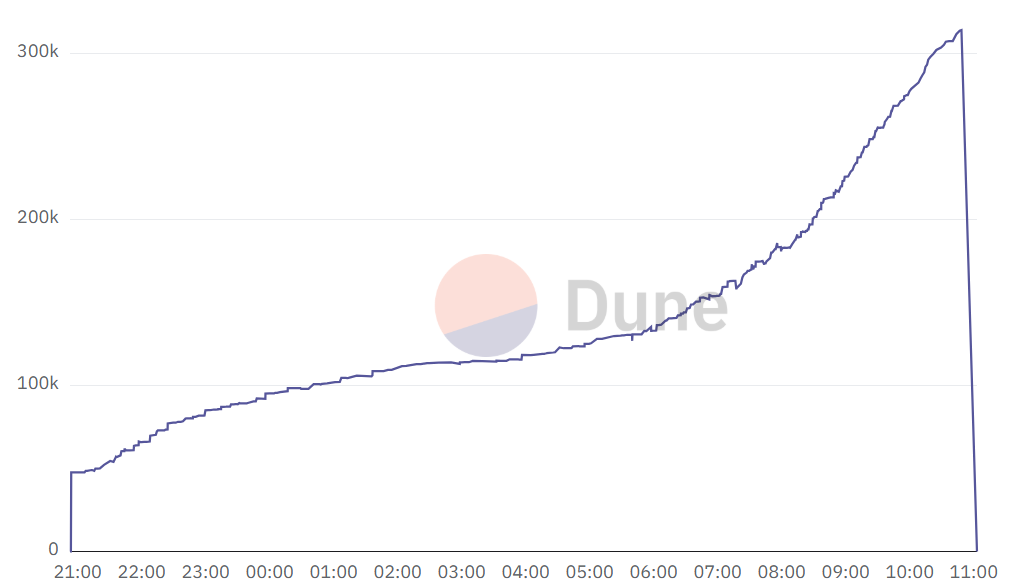}
    \caption{Example of TVL rug pull with \$UKWNPTHS token (chart shows TVL in USD since start of trading)}
    \label{fig:UKWNPTH}
\end{figure}

Figure \ref{fig:UKWNPTH} shows an example: \$UKWNPTHS token -- after reaching TVL peak, liquidity drops drastically, resulting in a loss of more than 99\% of its maximum value. TVL drops from \$300k to \$1 in a short period of time and activity ends. Formally, the method is defined by the “Maximum Drop in TVL”, the rug pull token $\Leftrightarrow MD \leq p$.

This approach mirrors the definition of a rug pull in that it relies directly on locked liquidity.

\begin{equation} 
t_0 = \underset
{t \in [0, 60]}{\arg\max}\ TVL(t)
\end{equation}
\begin{equation} 
\tau = \underset{t \in [t_0, 60]}{\arg\min}\ TVL(t)
\end{equation}
\begin{equation} 
MD = \dfrac{|TVL(t_0) - TVL(\tau)|}{TVL(t_0)}
\label{eq:MD}
\end{equation}

\subsection{Criteria for Model Evaluation}
\label{subsection:Criteria for Model Evaluation}

The following metrics are used to objectively evaluate the quality of the model:
\begin{itemize}
\item AUC (area under the ROC curve):
This metric mainly reflects the model's ability to distinguish between rug pulls and normal tokens.

\item Accuracy:
The proportion of correctly predicted rug pulls among all positive predictions. This is critical to minimizing false positives.

\item Recall:
The proportion of correctly predicted rug pulls among all true rug pulls. This is critical to minimize missing values (false negatives).

\item F1 score:
The harmonic mean of precision and recall, used to balance precision and recall.

\item Accuracy:
The overall proportion of correct predictions. However, due to class imbalance (which is common in rug pull tasks), this metric has less information.
\end{itemize}

This study will use the AUC metric to train the model and will use AUC, precision, and recall to analyze the results. The study will also consider the accuracy of Class 0 separately as it is crucial in the task of ensuring investor safety.

\subsection{Data Fusion}
\label{subsection:Data Fusion}
\begin{figure*}
    \centering
    \includegraphics[width=1\linewidth]{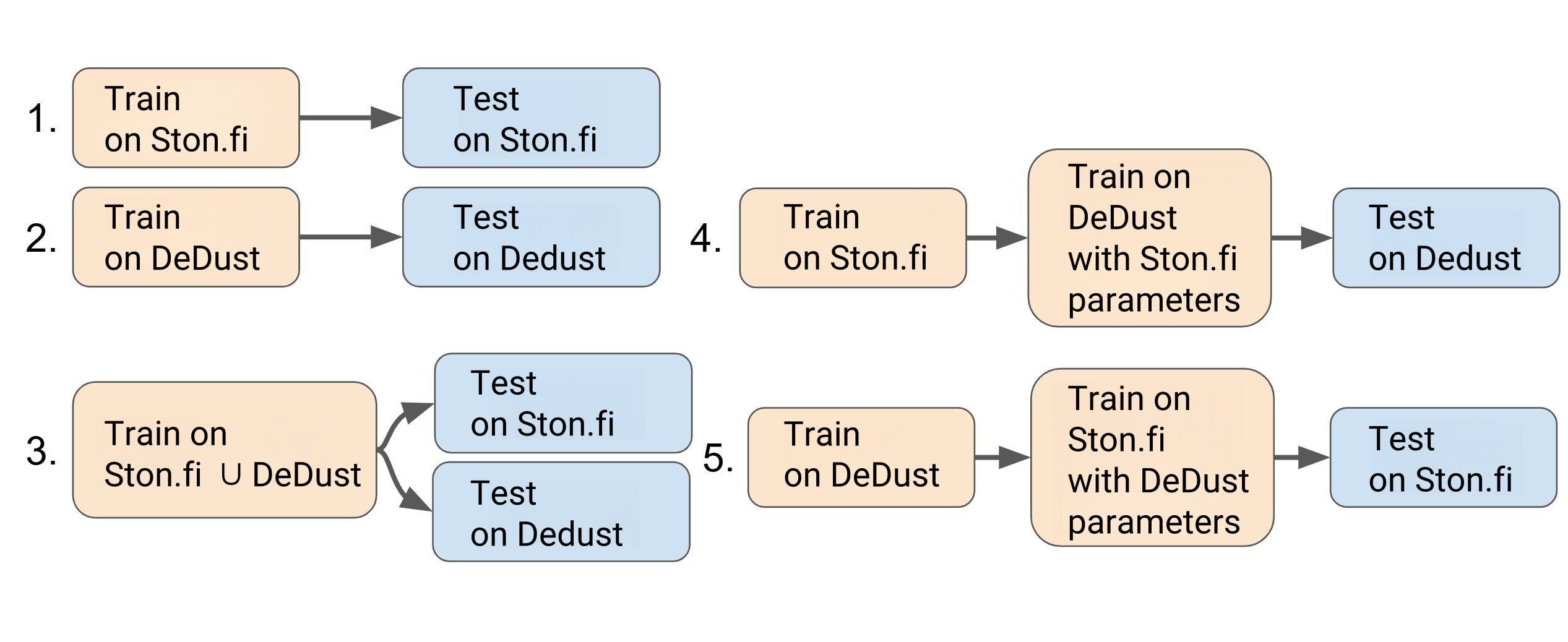}
    \caption{The process of data fusion on $ston.fi$ and $DeDust$ data}
    \label{fig:exps}
\end{figure*}

Model training and validation were performed not only on two samples (data sources), but also on various Data Fusion methods. Figure \ref{fig:exps} we can see five different approaches to model training:
\begin{enumerate}
\item Training on the $Ston.Fi$ sample, quality measurement on the same $Ston.Fi$ sample
\item Training on the $DeDust$ sample, quality measurement on the same $DeDust$ sample
\item Training on the combined $DeDust sample \;\cup\; Ston.Fi$, quality measurement on $Ston.Fi$, or on $DeDust$
\item Training on $Ston.Fi$ sample, saving model weights, then retraining on $DeDust$ sample, quality measurement on $DeDust$
\item Training on $DeDust$ sample, saving model weights, then retraining on $Ston.Fi$ sample, quality measurement on $Ston.Fi$
\end{enumerate}

The paper compares all methods to achieve a higher AUC.

\section{Data Mining}
\label{section:Data Mining}

\subsection{Data Collection}
\label{subsection:Data Collection}

To collect and analyze information on transactions, liquidity pools and trading history, this study used the dune.com service \cite{ton-dune}, which provides access to indexed blockchain data.

The TON Foundation, the organization responsible for developing and supporting the TON ecosystem, has played a special role in providing access to structured data. All transactions and events on the TON blockchain are indexed and aggregated in an open-source database that can be queried through dune.com. This allows researchers and developers to obtain up-to-date details on token transactions, the formation of liquidity pools, and other key events in the ecosystem.

To collect data, this study developed specialized SQL queries for ton.dex-trades, ton.dex-pools, and other tables in $dune.com$ to ensure maximum detail and relevance of the collected information.

\begin{figure}[!h]
    \centering
    \includegraphics[width=0.95\linewidth]{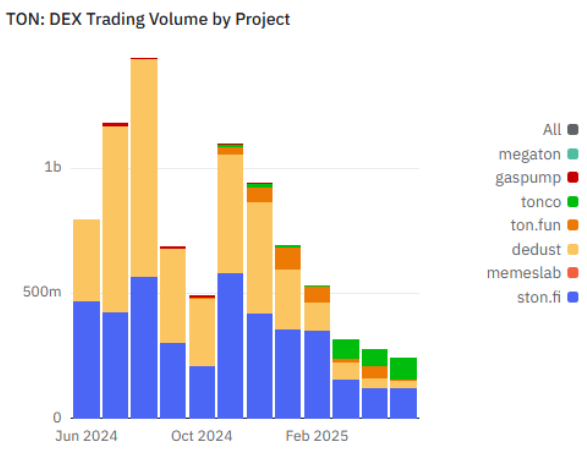}
    \caption{Trading volume of all DEX on TON for 2024-2025}
    \label{fig:dex_volume}
\end{figure}

Figure \ref{fig:dex_volume} shows that $Ston.Fi$ and $DeDust$ were taken for the study as the most representative and popular DEXs in TON.

Several DEXs were represented in the data.
\begin{itemize}
\item $Ston.Fi$ and $DeDust$ are top 1 and top 2 DEX in TON by volume of trading and popularity.
\item gaspump is a launchpad where tokens are sold via the Bonding Curve algorithm, and a pool is created automatically on $DeDust$ when liquidity reaches 1000 TON. No liquidity data, TVL approach cannot be used.
\item ton.fun is a pump.fun analogue on TON. No liquidity data, TVL approach cannot be used.
\item tonco has data on only 80 tokens.
\item megaton has data on only 38 tokens.
\item memeslab had no data at the time of model training and data collection.
\end{itemize}

\begin{figure*}[!h]
    \centering
    \includegraphics[width=0.95\linewidth]{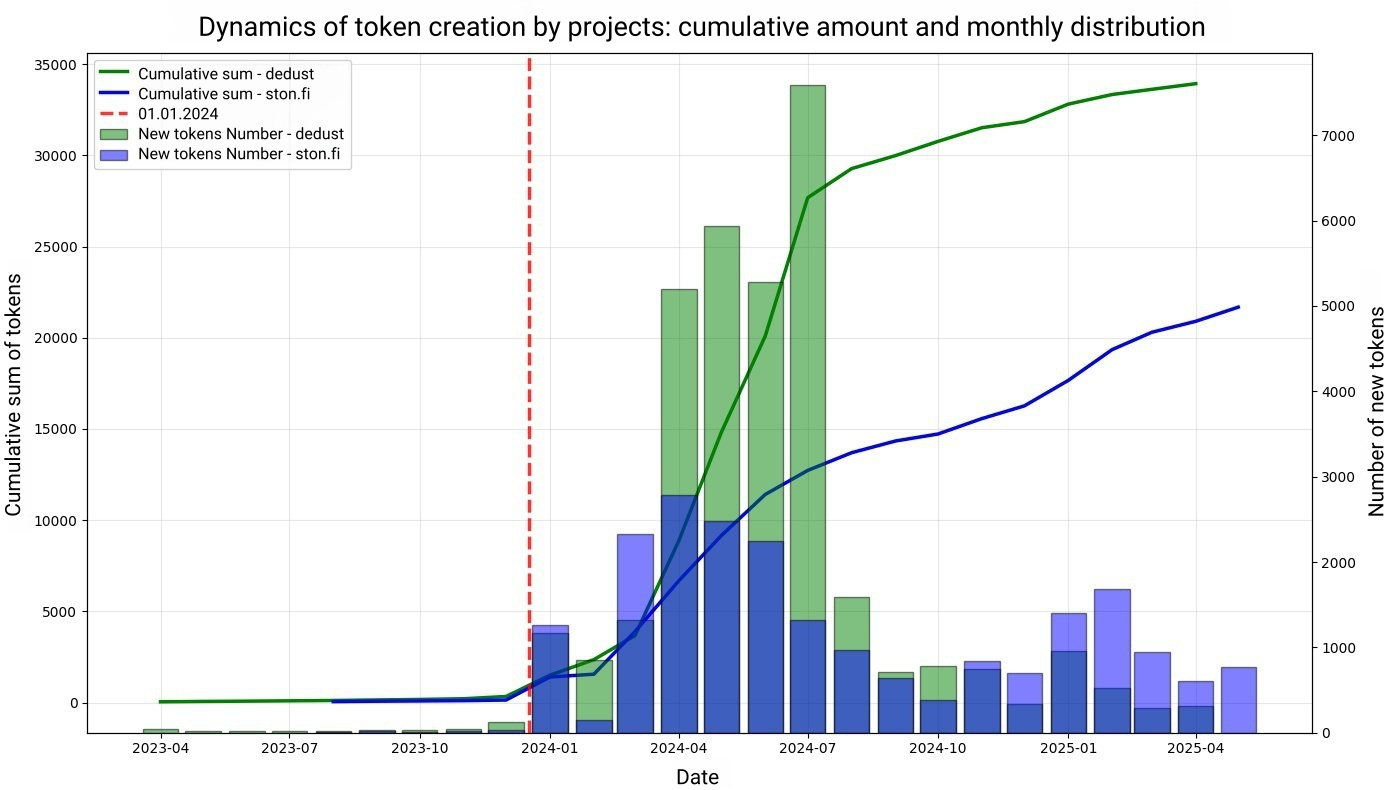}
    \caption{Per month number of new tokens created on $Ston.Fi$ and $DeDust$}
    \label{fig:dex_tokens}
\end{figure*}

Figure \ref{fig:dex_tokens} shows that the lower time limit of Jan 1, 2024 was also chosen as tokens with a trading start date no earlier than Jan 1, 2024 and no later than Apr 1, 2025 account for 99.4\% of all tokens (for $Ston.Fi$ – 99.4\%, for $DeDust$ – 99.1\%). Information on the token ratio was collected in May 2025.

\subsection{Dataset Description}
\label{subsection:Dataset Description}

The samples contain information on each token, a detailed description of the features can be found in the Table \ref{tab:features}.

The samples included only tokens with pool information (there was at least one record about the pool and its liquidity), only transactions with a non-zero value of $volume\_usd$, since we are only interested in token purchases and sales, and only pools from the TON blockchain.

As a result, the sample sizes were: 30,097 tokens on $DeDust$ and 18,283 tokens for $Ston.Fi$.

\begin{table*}[!h]
\centering
\begin{tabularx}{\linewidth}{cZ}
\toprule
Feature & Description \\
\midrule

buy\_sell\_ratio & Ratio of token purchases and sales \\
price\_range & Difference between maximum and minimum prices \\
buys & Number of purchases in 5 minutes \\
sells & Number of sales in 5 minutes \\
buy\_perc & Percentage of purchases from all transactions \\
sell\_perc & Percentage of sales from all transactions \\
unique\_buyers & Number of unique buyers \\
unique\_sellers & Number of unique sellers \\
total\_usd\_volume & Total transaction volume in dollars \\
total\_usd\_buy\_volume & Purchase volume in dollars \\
total\_usd\_sell\_volume & Sales volume in dollars \\
decimals & Technical field for token (precision) \\
avg\_lp\_fee & Average LP fee \\
avg\_protocol\_fee & Average protocol fee \\
jetton\_creation\_trade\_delta & Difference between token sale start and token creation (in seconds) \\
pool\_creation\_trade\_delta & Difference between token sale start and pool creation (in seconds) \\
is\_pool\_creator & Token creator is the same as pool creator \\
initial\_tvl\_usd & Initial TVL in pool \\
initial\_price & Token price in first transaction \\
initial\_buy\_price & Token price in first purchase \\
max\_tvl & Maximum TVL in 5 minutes \\
min\_tvl & Minimum TVL in 5 minutes \\
buy\_price\_std & Standard deviation of purchase price \\
initial\_sell\_price & Price token in first sale \\
sell\_price\_std & Standard deviation of sale price \\
price\_max & Maximum price in 5 minutes \\
price\_min & Minimum price in 5 minutes \\
price\_delta & Difference between last and starting price \\
price\_std & Standard deviation of price in transactions \\
first\_buy\_time\_ts & Time of first purchase \\
first\_sell\_time\_ts & Time of first sale \\
pool\_deployment\_at\_ts & Pool deployment time \\
jetton\_deployment\_at\_ts & Token deployment time \\

\bottomrule
\end{tabularx}
\caption{Description of feature in samples}
\label{tab:features}
\end{table*}

\subsection{Feature Engineering}
\label{subsection:Feature Engineering}

Distribution histograms were constructed for all numerical features. Most features have a right-skewed distribution, which is typical for financial data: most tokens have low activity, and a small number demonstrate high values in terms of volume, number of transactions, and TVL.

\subsection{Data Labeling}
\label{subsection:Data Labeling}

In the definition of TVL rug pull there is a variable $p$ ~---~ acceptable percentage of TVL drop.

\begin{figure}[!h]
    
    \centering
    \includegraphics[width=0.95\linewidth]{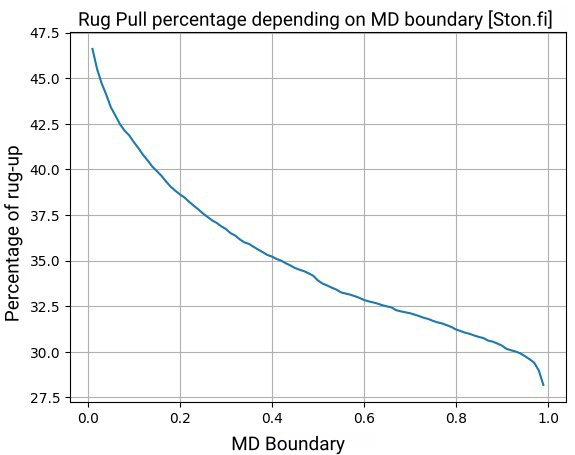}
    \caption{Rug percentage as a function of p for Ston.Fi}
    \label{fig:md_rug_stonfi}

    \centering
    \includegraphics[width=0.95\linewidth]{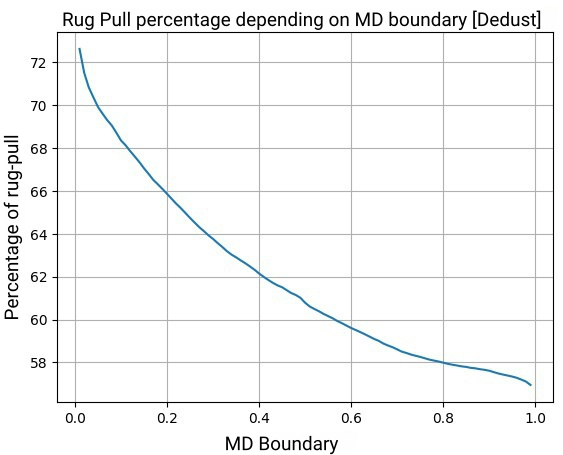}
    \caption{Rug percentage as a function of p for DeDust}
    \label{fig:md_rug_dedust}
    
\end{figure}

In Figure \ref{fig:md_rug_stonfi} and Figure \ref{fig:md_rug_dedust}, it can see graphs of the dependence of the rug pull percentage for the TVL approach on the acceptable percentage of $MD$ ~---~ $p$ drop, for $Ston.Fi$ and $DeDust$ respectively \cite{mazorra2022rug}.

It is also necessary to determine how much time ahead this study will predict rug pull, similar to how the distribution of the ratio of classes 0 and 1 was constructed depending on how long it predicted rug pull \cite{rugpulls2024}. It is clear that the classes in both the TVL and Idle approaches are not balanced. This is important to consider when training models to avoid overfitting to the majority class. For this work, a time of 60 minutes was chosen.

\begin{figure*}[!h]
    \centering
    \includegraphics[width=0.8\linewidth]{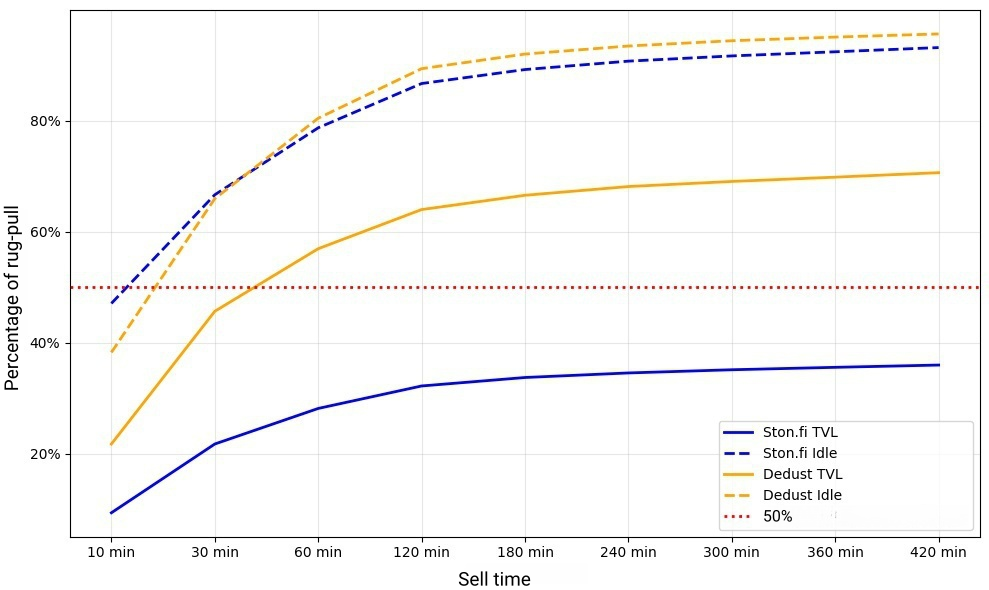}
    \caption{Dependence of the ratio of 0 and 1 classes on the prediction horizon of rug pull}
    \label{fig: when_sell}
\end{figure*}

\subsection{Data Preprocessing Techniques} 
\label{subsection:Data_Preprocessing_Techniques}

During the preliminary data analysis, problems related to the quality of the datasets were identified: the presence of zero values, NULL, and noise in the form of outliers. These features are typical for data collected from decentralized exchanges and require special processing.

\begin{itemize}
\item \textbf{NULLs:} Some features, especially metadata and timestamps, contain missing values. To eliminate them, they were filled with median values for numeric features.
\item \textbf{Null values:} Often found in features related to transaction volume (often token sales). Such values may indicate low token activity.
\item \textbf{Noise:} The data contain outliers in volume and numbers of transactions. To reduce the impact of noise, scaling and algorithmization methods were used.
\end{itemize}

\begin{figure}[h!]
\centering
\begin{subfigure}{0.49\textwidth}
    \centering
    \includegraphics[width=\linewidth]{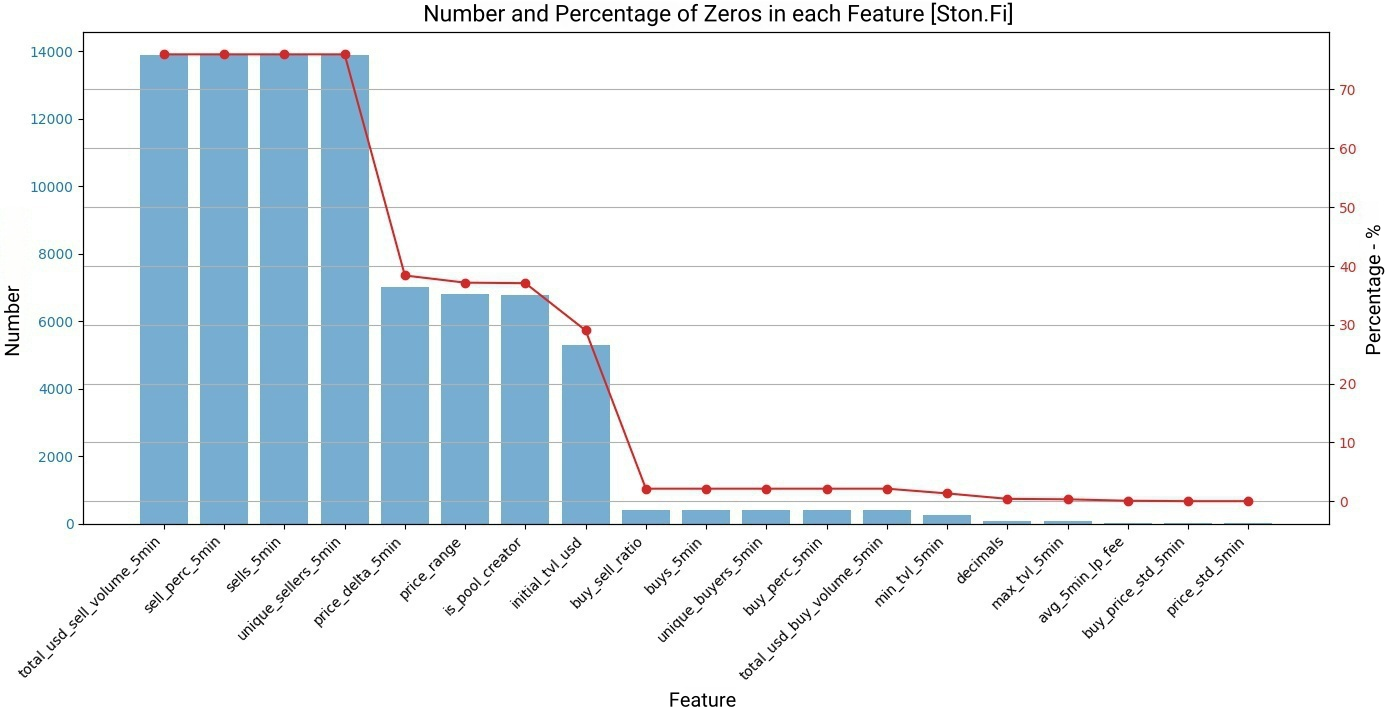}
    \caption{Distribution of Zero values ($Ston.Fi$)}
    \label{fig:zeros_stonfi}
\end{subfigure}
\hfill
\begin{subfigure}{0.49\textwidth}
    \centering
    \includegraphics[width=\linewidth]{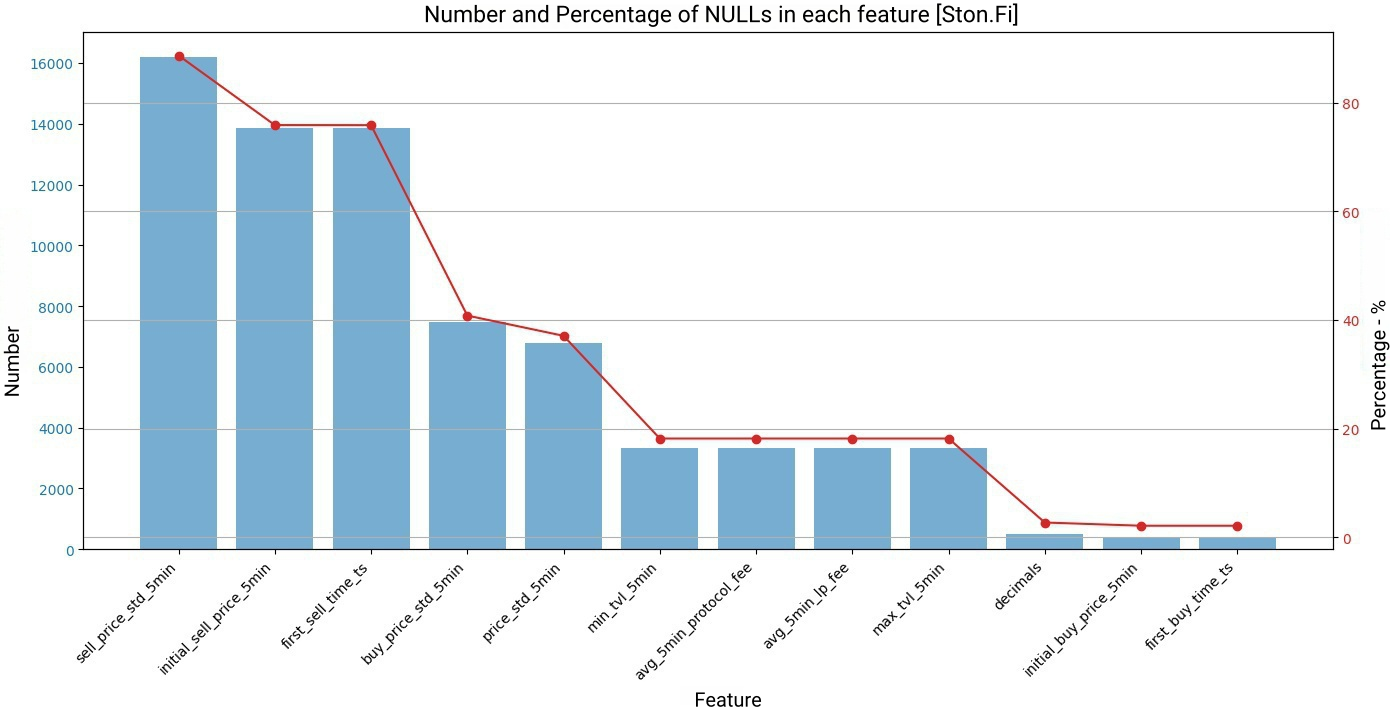}
    \caption{Distribution of NULL values ($Ston.Fi$)}
    \label{fig:nulls_stonfi}
\end{subfigure}

\vspace{0.5cm}

\begin{subfigure}{0.49\textwidth}
    \centering
    \includegraphics[width=\linewidth]{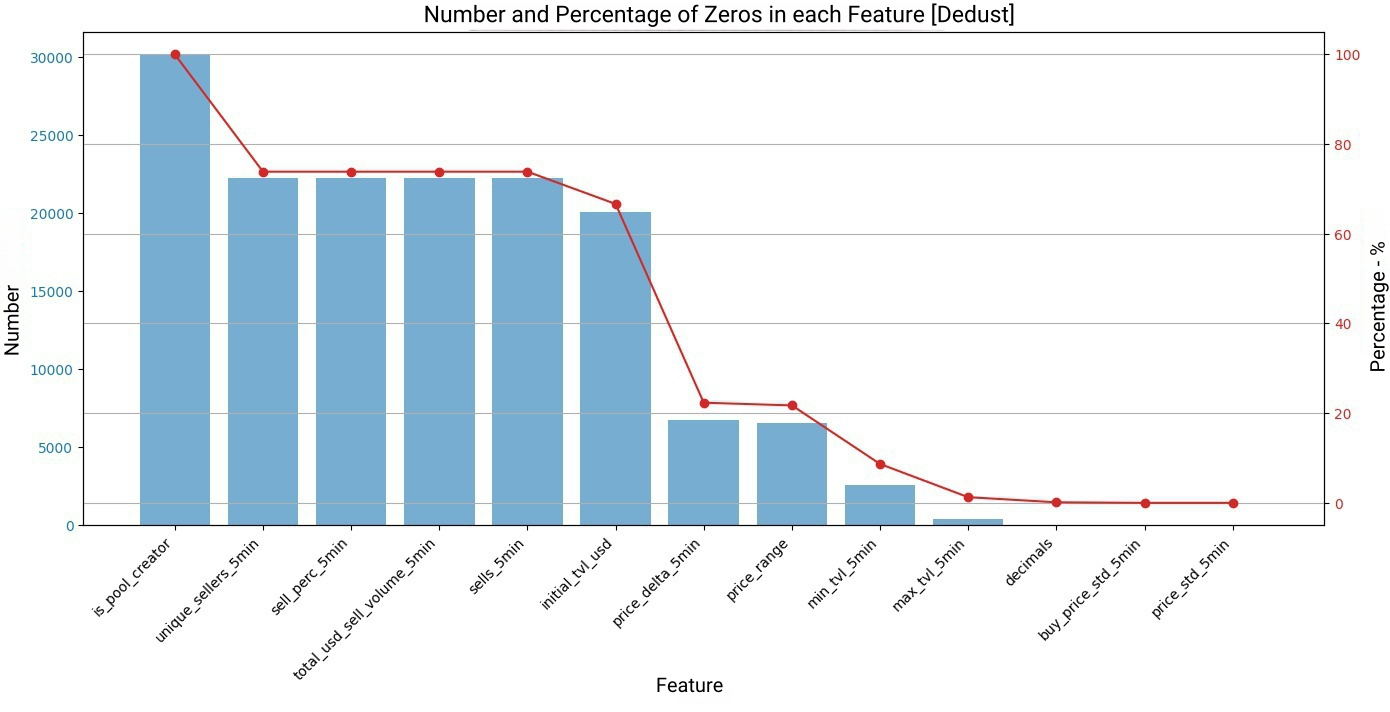}
    \caption{Distribution of Zero values ($DeDust$)}
    \label{fig:zeros_DeDust}
\end{subfigure}
\hfill
\begin{subfigure}{0.49\textwidth}
    \centering
    \includegraphics[width=\linewidth]{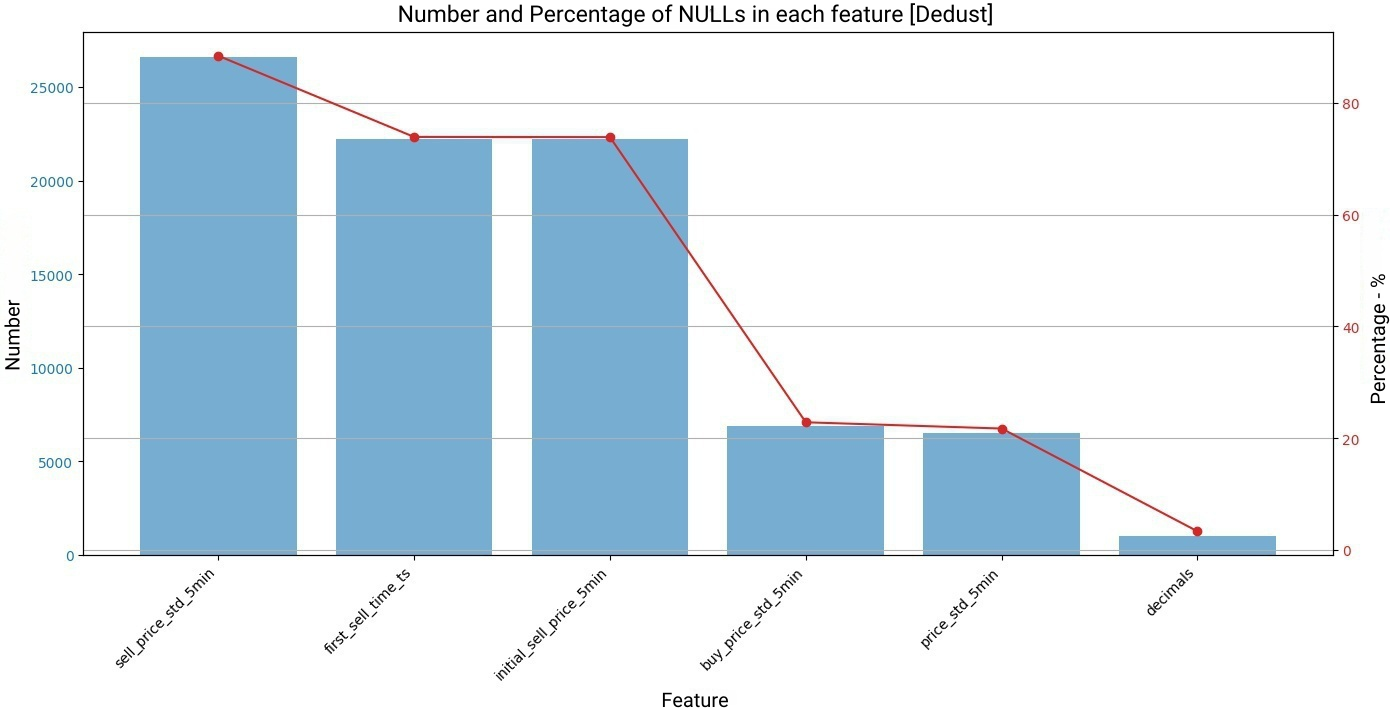}
    \caption{Distribution of NULL values ($DeDust$)}
    \label{fig:nulls_DeDust}
\end{subfigure}

\caption{Comparative analysis of the distribution of Zero and NULL values in the $Ston.Fi$ and $DeDust$}
\label{fig:nulls_zeros_comparison}
\end{figure}

Figure \ref{fig:nulls_zeros_comparison} shows the visualization of the distribution of zero and NULL features. NULLs and zeros are mostly due to missing token sales, which are about 80\% in both samples, and also due to missing metadata such as the token creator's address or $decimals$.

\section{Detection Model}
\label{section:Detection Model}

\subsection{Data Preprocessing and Feature Engineering}
\label{subsection:Data Preprocessing and Feature Engineering}

For each sample ($Ston.Fi$ and $DeDust$), the following steps were performed:

\begin{itemize}
\item \textbf{Gap removal:} Rows with missing values in key features (e.g. timestamps, TVL) were excluded from the sample.
\item \textbf{Zero filling:} For numeric features with zero values, median filling or row removal were used.
\item \textbf{Scaling:} Numeric features were scaled using StandardScaler (or left unscaled depending on the results of cross-validation).
\item \textbf{Feature generation:} Additional features were created, such as the buy/sell ratio, price range, standard price deviation, time deltas between token creation and the first transaction, etc.
\item \textbf{Time feature processing:} Timestamps were converted to numeric deltas (e.g., the difference between token creation and start of trading in seconds). $timestamp$ time features were also used as features.
\end{itemize}

\subsection{Model and Hyperparameter Tuning}
\label{subsection:Model and Hyperparameter Tuning}

\subsubsection{Machine Learning Model}

The following models were used to solve the rug pull binary classification problem:

\begin{itemize}
\item \textbf{GradientBoosting} -- boosting algorithm with sequential tree construction.
\item \textbf{RandomForest} -- ensemble of decision trees with bagging.
\item \textbf{DecisionTree} -- basic decision tree algorithm.
\item \textbf{ExtremeGradientBoosting} -- gradient boosting of trees with regularization, effective for problems with class imbalance.
\end{itemize}

The choice of models is due to their high efficiency in solving class imbalance problems and the ability to interpret the importance of features.

\subsubsection{Hyperparameter tuning}

For each model, hyperparameters were selected using GridSearchCV:
\begin{itemize}
\item \textbf{XGBoost:} learning\_rate, max\_depth, n\_estimators, subsample, colsample\_bytree, gamma, reg\_alpha, reg\_lambda.
\item \textbf{GradientBoosting:} learning\_rate, max\_depth, n\_estimators, subsample, loss, min\_samples\_split, max\_features.
\item \textbf{RandomForest:} n\_estimators, max\_depth, min\_samples\_split, max\_features, class\_weight.
\item \textbf{DecisionTree:} max\_depth, min\_samples\_split, max\_features.
\end{itemize}

The selection was carried out on cross-validation with splitting into 3 folds to minimize overfitting and maximize the quality of the models. Also, for each model, the data preprocessing method was tried, either StandardScaler or leave as is, and this was done separately for numerical and temporal features.

\subsubsection{Cross-validation}

To assess the quality of the models, a division into training and test samples was used in a ratio of 8:2, while maintaining the proportions of classes. At the same time, when training on a combined data set ($DeDust \;\cup\; Ston.Fi$), stratification was performed not only by class but also by data source.
Training sample for 3 folds, maintaining the balance of classes in each fold.

\section{Numerical Results}
\label{section:Numerical_Results}

\subsection{Comparison of Model Performance}
\label{subsection:Comparison of model performance}

\subsubsection{AUC Analysis}

During the experiments, the following results were obtained for the AUC metric for the best models in each dataset and approache: Idle in Table \ref{models1} and Table \ref{models2}, TVL in Table \ref{models3} and Table \ref{models4}.
The notation $DeDust\;\cup\; Ston.Fi$ means training in the combined sample, and measuring the quality only in $DeDust$ or $Ston.Fi$, and $Ston.Fi \rightarrow DeDust$ means training and measuring in the data $DeDust$, but for a model whose weights are preserved after training in the sample $Ston.Fi$.

\begin{table*}[!h]
\centering
\scriptsize
\captionsetup{justification=centering}
\begin{tabularx}{\linewidth}{ZZZZ}
\toprule

Metrics / Experiment & $DeDust$ & $DeDust\; \cup\; Ston.Fi$ & $Ston.Fi \rightarrow DeDust$ \\

\textbf{Best Model} & ExtremeGradientBoosting & ExtremeGradientBoosting & ExtremeGradientBoosting \\

\midrule
\textbf{Precision (1)} & 0.86 & 0.86 & 0.90 \\
\textbf{Recall (1)} & 0.96 & 0.97 & 0.82 \\
\textbf{F1 (1)} & 0.91 & 0.91 & 0.86 \\
\textbf{Precision (0)} & 0.70 & \textbf{0.78} & 0.46 \\
\textbf{Recall (0)} & 0.35 & 0.36 & 0.64 \\
\textbf{F1 (0)} & 0.46 & 0.49 & 0.54 \\
\textbf{AUC} &\textbf{0.820} & \textbf{0.820} & \textbf{0.820} \\

\bottomrule
\end{tabularx}
\caption{Experimental results for the \textbf{Idle} approach on \textbf{DeDust} data}
\label{models1}
\end{table*}

\begin{table*}[!h]
\centering
\scriptsize
\captionsetup{justification=centering}
\begin{tabularx}{\linewidth}{ZZZZ}
\toprule
Metrics / Experiment & $Ston.Fi$ & $DeDust\; \cup\; Ston.Fi$ & $DeDust \rightarrow Ston.Fi$ \\ 

\textbf{Best Model} & ExtremeGradientBoosting & RandomForest & ExtremeGradientBoosting \\

\midrule

\textbf{Precision (1)} & 0.87 & 0.87 & 0.85 \\
\textbf{Recall (1)} & 0.97 & 0.96 & 0.99 \\
\textbf{F1 (1)} & 0.92 & 0.91 & 0.92 \\
\textbf{Precision (0)} & 0.79 & 0.72 & \textbf{0.90} \\
\textbf{Recall (0)} & 0.47 & 0.41 & 0.37 \\
\textbf{F1 (0)} & 0.58 & 0.52 & 0.52 \\
\textbf{AUC} & \textbf{0.840} & 0.838 & \textbf{0.840} \\

\bottomrule
\end{tabularx}
\caption{Experimental results for the \textbf{Idle} approach on \textbf{Ston.Fi} data}
\label{models2}
\end{table*}

\begin{table*}[!h]
\centering
\scriptsize
\captionsetup{justification=centering}
\begin{tabularx}{\linewidth}{ZZZZ}
\toprule
Metrics / Experiment & $Ston.Fi$ & $DeDust\; \cup\; Ston.Fi$ & $DeDust \rightarrow Ston.Fi$ \\ 

\textbf{Best Model} & GradientBoosting  & ExtremeGradientBoosting  & ExtremeGradientBoosting \\

\midrule

\textbf{Precision (1)} & 0.76 & 0.74 & 0.74 \\
\textbf{Recall (1)} & 0.64  & 0.81 & 0.65 \\
\textbf{F1 (1)} & 0.69  & 0.78 & 0.69 \\
\textbf{Precision (0)} & \textbf{0.87}  & 0.83 & \textbf{0.87} \\
\textbf{Recall (0)} & 0.92  & 0.76 & 0.91 \\
\textbf{F1 (0)} & \textbf{0.89}  & 0.79 & \textbf{0.89} \\
\textbf{AUC} & 0.885 & \textbf{0.891} & 0.885 \\

\bottomrule
\end{tabularx}
\caption{Experimental results for the \textbf{TVL} approach on \textbf{Ston.Fi} data}
\label{models3}
\end{table*}

\begin{table*}[!h]
\centering
\scriptsize
\captionsetup{justification=centering}
\begin{tabularx}{\linewidth}{ZZZZ}
\toprule

Metrics / Experiment & $DeDust$ & $DeDust\; \cup\; Ston.Fi$ & $Ston.Fi \rightarrow DeDust$ \\

\textbf{Best Model} & GradientBoosting & GradientBoosting  & ExtremeGradientBoosting  \\

\midrule

\textbf{Precision (1)} & 0.76  & 0.77  & 0.76  \\
\textbf{Recall (1)} & 0.82  & 0.78 &  0.81  \\
\textbf{F1 (1)} & 0.79  & 0.77 &  0.78  \\
\textbf{Precision (0)} & 0.73 & \textbf{0.81} &  0.72  \\
\textbf{Recall (0)} & 0.67  & 0.80  & 0.67 \\
\textbf{F1 (0)} & 0.70 & 0.80 &  0.69  \\
\textbf{AUC} & 0.829 & \textbf{0.831} & 0.826 \\

\bottomrule
\end{tabularx}
\caption{Experimental results for the \textbf{TVL} approach on \textbf{DeDust} data}
\label{models4}
\end{table*}

\begin{figure}
\centering
\includegraphics[width=1\linewidth]{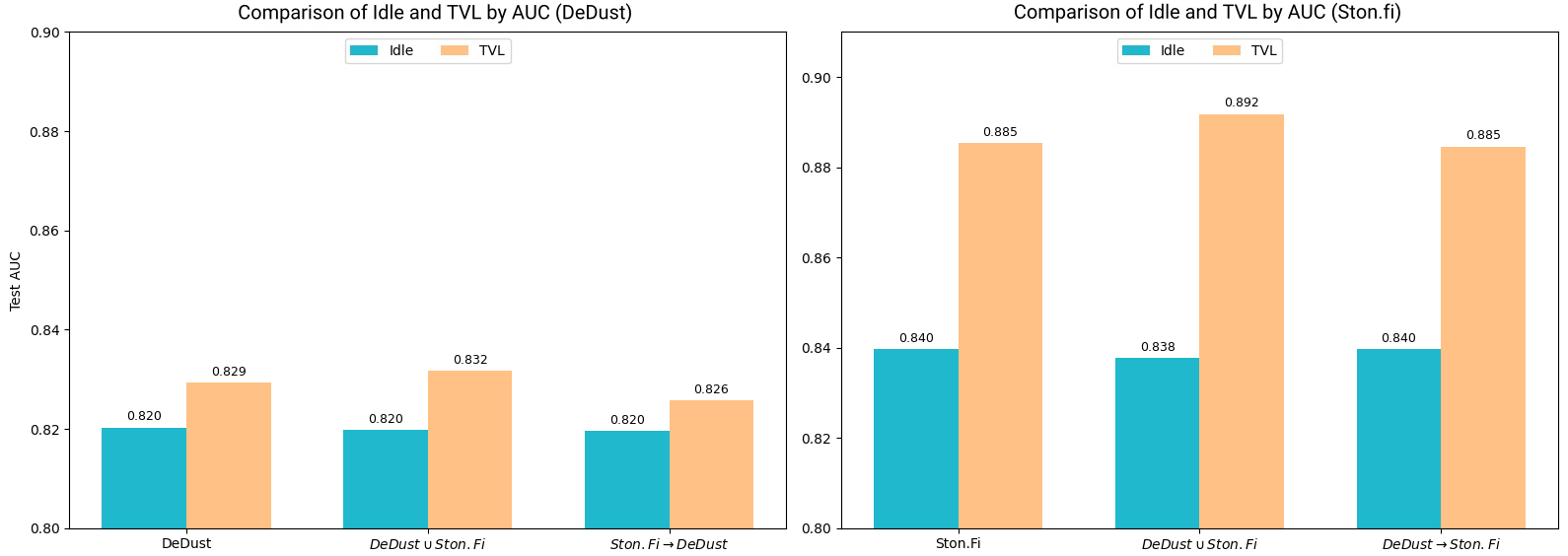}
\caption{AUC results for different approaches and data sources}
\label{fig:idle_vs_tvl_auc_new}
\end{figure}

For clarity, a comparison chart of AUC for different datasets and approaches is provided below Figure \ref{fig:idle_vs_tvl_auc_new}.

The analysis of the results shows that the TVL approach shows a higher average AUC value ($0.860$) compared to the Idle approach ($0.829$), with a difference of $0.031$.

Mixed learning on combined data ($DeDust\; \cup\; Ston.Fi$) does not improve the AUC metric for $DeDust$, but positively affects $Ston.Fi$ TVL approach ($+0.007$)

Also, for the TVL approach, testing in $Ston.Fi$ data shows significantly better results with an average AUC of $0.887$, versus $0.829$ on $DeDust$ data. This may indicate better data quality and clearer rug pull patterns on the $Ston.Fi$ platform.

Applying Recursive Feature Elimination (RFE) resulted in a slight improvement in the metrics, achieving an accuracy of 87.89\%. This algorithm uses a recursive iterative process, each time training on a subset of features, selecting the least important ones for the model, and eliminating them. In this case, the algorithm identified the following features for elimination: creation\_month\_cos, is\_pool\_creator, and std\_rsi. 

\subsubsection{Comparison of Accuracy Class 0}

\begin{figure}
\centering
\includegraphics[width=1\linewidth]{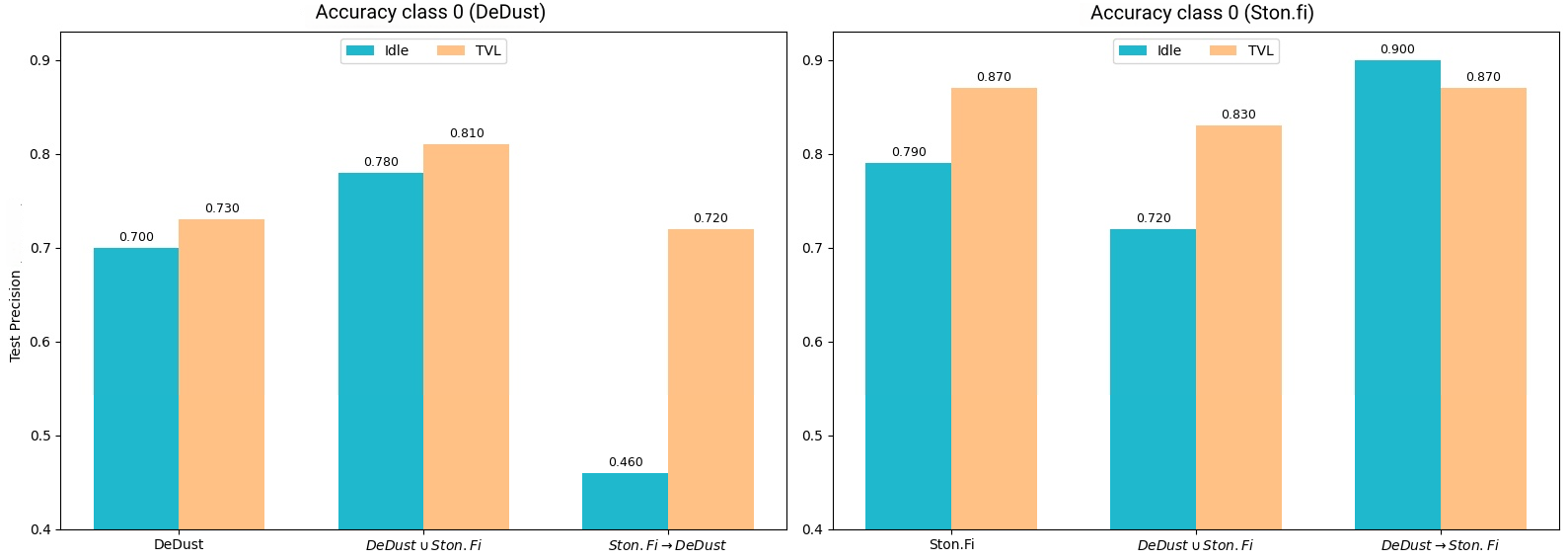}
\caption{Class 0 accuracy results for different approaches and data sources}
\label{fig:idle_vs_tvl_prec_0_new}
\end{figure}

For clarity, a Precision (0) comparison chart for different datasets and approaches is provided in Figure \ref{fig:idle_vs_tvl_prec_0_new}.

The accuracy of detecting non-rug pull tokens (class 0) is a critical metric, as false positives reduce user confidence in the warning system.

The best Precision(0) is achieved using the Idle approach with transfer learning from $DeDust$ to $Ston.Fi$ (0.90)

With the idle approach, transferring weights from $DeDust$ to $Ston.Fi$ improves precision (0) to 0.90, while transferring from $Ston.Fi$ to $DeDust$ reduces it to 0.46. For the TVL approach, transfer learning shows stable results in both directions.

\subsection{Feature Importance Analysis}
\label{subsection:Feature Importance Analysis}

Feature importance analysis was performed for the best models in each sample. In the Idle approach, the most significant features were those related to transaction volume and number of purchases (e.g. \texttt{total\_usd\_volume\_5min}, \texttt{buys\_5min}, \texttt{is\_pool\_creator} for $Ston.Fi$ and $DeDust$). In the TVL approach, the key features are related to liquidity and time deltas (e.g. \texttt{max\_tvl\_5min}, \texttt{jetton\_creation\_trade\_delta}, \texttt{first\_buy\_time\_ts} for $Ston.Fi$).

\begin{figure}
\includegraphics[width=0.8\linewidth]{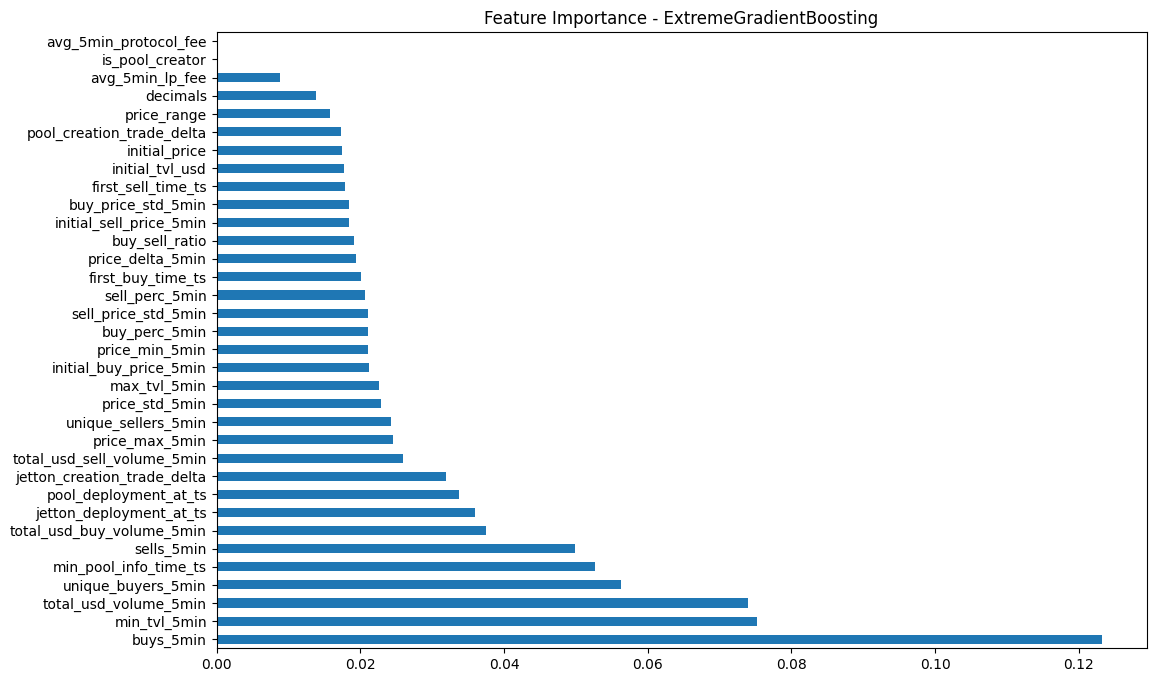}
\caption{Feature importance for the Idle approach on $DeDust$ data}
\label{fig:DeDust_best_idle}
\end{figure}

\begin{figure}
\includegraphics[width=0.8\linewidth]{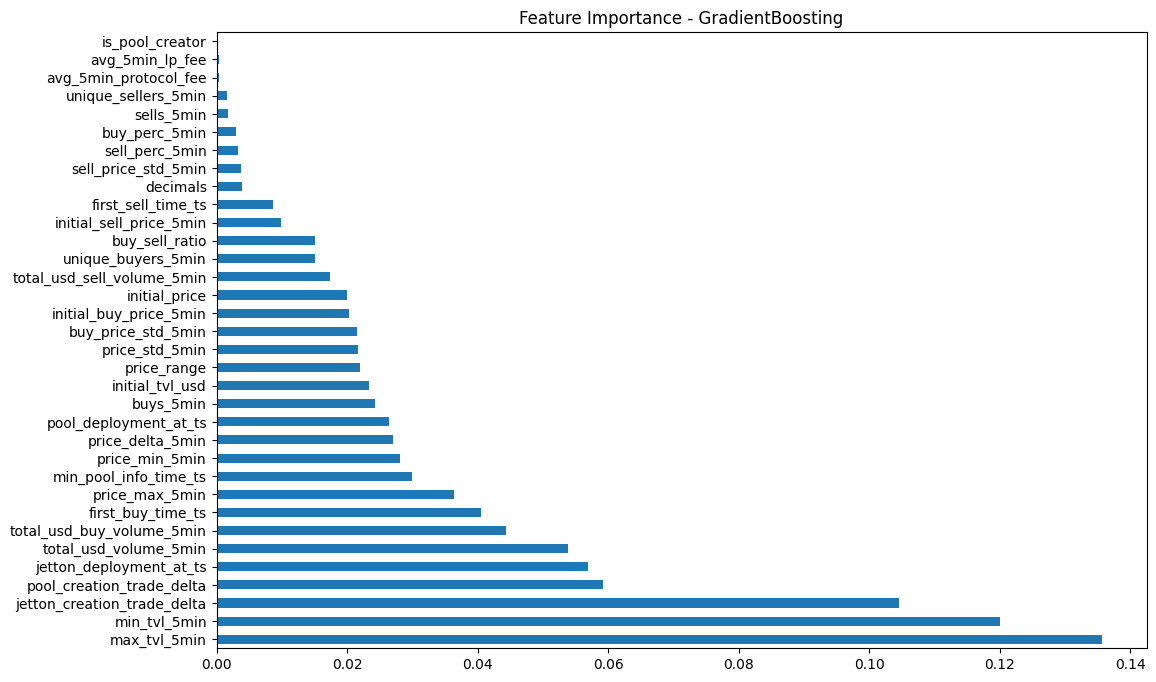}
\caption{Feature importance for TVL approach on $DeDust$ data}
\label{fig:DeDust_best_tvl}
\end{figure}

\begin{figure}
\includegraphics[width=0.8\linewidth]{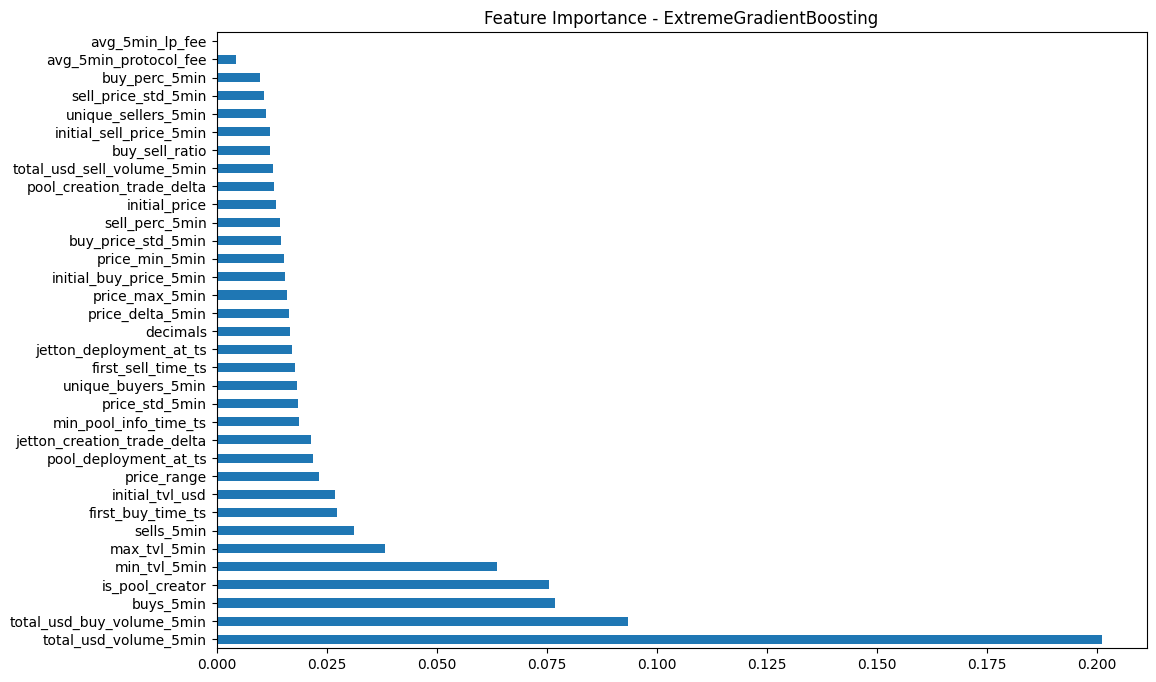}
\caption{Feature importance for Idle approach on $Ston.Fi$ data}
\label{fig:stonfi_best_idle}
\end{figure}

\begin{figure}
\includegraphics[width=0.8\linewidth]{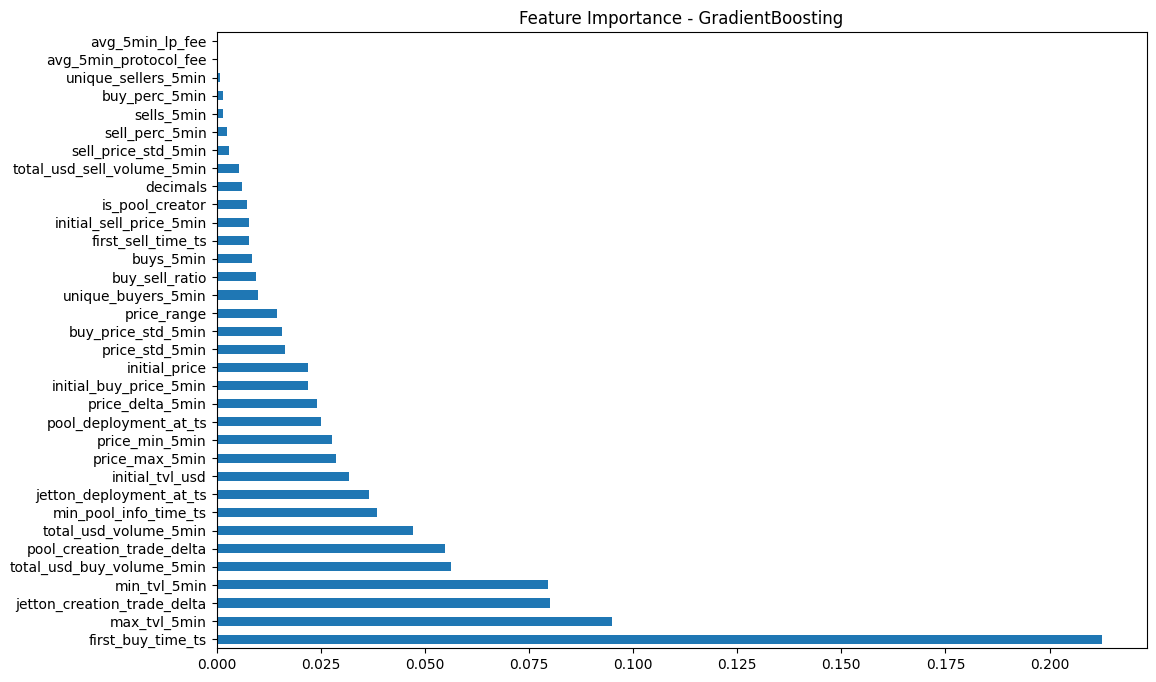}
\caption{Feature importance for TVL approach on $Ston.Fi$ data} 
\label{fig:stonfi_best_tvl}
\end{figure}

For clarity, visualizations of feature importance for each approach (Idle, TVL) and each data source ($DeDust$, $Ston.Fi$) are presented below Figure \ref{fig:DeDust_best_idle} \ref{fig:DeDust_best_tvl} \ref{fig:stonfi_best_idle} \ref{fig:stonfi_best_tvl}.

\subsection{Comparison of Idle and TVL approaches} 
\label{subsection:Comparison of Idle and TVL approaches}

Comparison of the Idle and TVL approaches showed that the TVL approach demonstrates higher and more stable performance on both samples by AUC metric. This is due to the fact that TVL is a more objective and informative indicator of rug pull compared to no transactions (Idle), especially for tokens with low activity. However, the best Precision(0) indicator is achieved using the Idle approach with transfer learning from $DeDust$ to $Ston.Fi$ (0.90), making the Idle approach more applicable to real life.

The Idle approach better identifies class 1 (rug pull), and the TVL approach better identifies class 0 (not rug pull); this is clearly seen in the table with precision and recall by classes. This is natural because these approaches had a corresponding skew in the ratio of classes.

\subsection*{Experimental Summary}

The experiments confirmed the high efficiency of the proposed approach in detecting rug pull in DEX $Ston.Fi$ and $DeDust$. The best results were shown by the GradientBoosting and ExtremeGradientBoosting models, especially when using the TVL approach. Combining data from different exchanges (data fusion) allows to increase the stability and quality of the models. The key features for predicting rug pull are the volume of transactions, the number of purchases, and the liquidity of the token. The TVL approach is recommended as the main one for the practical implementation of the rug pull monitoring system on the TON blockchain.


\section{Discussions and Conclusions}
\label{section:Discussions and Conclusions}

\subsection*{Comparative analysis of results by datasets and approaches}

\begin{figure*}[ht]
\begin{subfigure}{0.48\linewidth} 
\includegraphics[width=\linewidth]{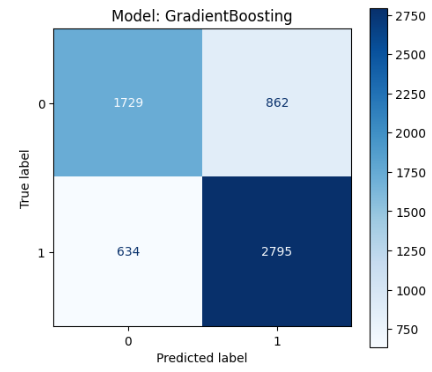} 
\caption{Based on data from $DeDust$} 
\label{fig:matrix_tvl_DeDust}
\end{subfigure}
\begin{subfigure}{0.48\linewidth} 
\includegraphics[width=\linewidth]{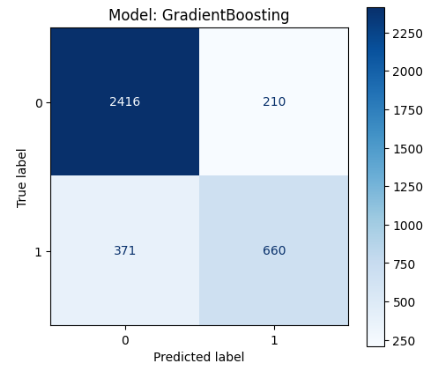} 
\caption{Based on data from $Ston.Fi$} 
\label{fig:matrix_tvl_stonfi}
\end{subfigure}
\caption{Error matrix for TVL approach}
\label{fig:matrix_tvl}
\end{figure*}

\begin{figure*}[ht]
\begin{subfigure}{0.48\linewidth}
\includegraphics[width=\linewidth]{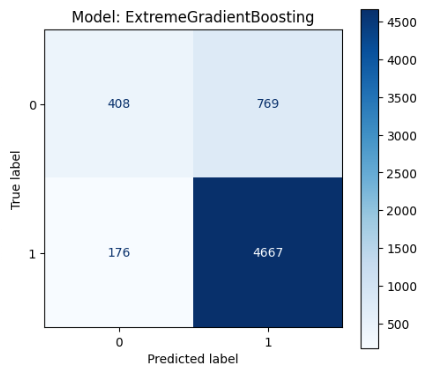}
\caption{Based on $DeDust$ data}
\label{fig:matrix_idle_DeDust}
\end{subfigure}
\begin{subfigure}{0.48\linewidth}
\includegraphics[width=\linewidth]{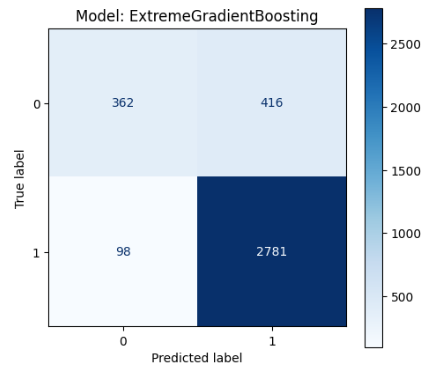}
\caption{Based on $Ston.Fi$ data}
\label{fig:matrix_idle_stonfi}
\end{subfigure}
\caption{Error matrix for Idle approach}
\label{fig:matrix_idle}
\end{figure*}

The experiments yielded results that allow this study to evaluate the effectiveness of various machine learning models to detect rug pull on the $Ston.Fi$ and $DeDust$ DEX platforms, as well as on the combined dataset (data fusion). The main conclusions for each approach and dataset are presented in the following.

\textbf{Idle approach} (rug pull - no transactions within an hour after the start of trading):
\begin{itemize}
\item On the $Ston.Fi$ dataset, the best results were shown by the \\ ExtremeGradientBoosting model: AUC = 0.8397, precision (1) = 0.87, recall (1) = 0.97, which indicates a high ability of the model to detect rug pull at the early stages of trading.
\item On the $DeDust$ dataset, ExtremeGradientBoosting also leads: AUC = 0.8203, precision (1) = 0.86, recall (1) = 0.96.
\item On the combined dataset ($Ston.Fi \;\cup\; DeDust$), the best model is \\ RandomForest: AUC = 0.8265, precision (1) = 0.87, recall (1) = 0.96.
\item Models trained on the combined dataset and tested on another demonstrate comparable quality, which confirms the robustness of the approaches to differences in data distributions between exchanges.
\end{itemize}

\textbf{TVL approach} (rug pull - TVL drop by more than 99\% from the maximum in the first hour of trading):
\begin{itemize}
\item In $Ston.Fi$, the best model is GradientBoosting: AUC = 0.8853, precision (1) = 0.76, recall (1) = 0.64.
\item On $DeDust$, GradientBoosting is also in the lead: AUC = 0.8293, precision (1) = 0.76, recall (1) = 0.82.
\item On the combined dataset, the best model is GradientBoosting: AUC = 0.8713, precision (1) = 0.77, recall (1) = 0.78.
\item Models trained on the combined dataset and tested on individual exchanges show high generalizability and stability of results.
\end{itemize}

A comparative analysis of the approaches showed that the TVL approach demonstrates higher AUC values in $Ston.Fi$, and the Idle approach shows higher recall values in both data sets. This is due to the specifics of rug pull detection: the TVL approach is more sensitive to a sharp drain on liquidity, and the Idle approach to the cessation of activity.

\subsection*{Practical Application Suggestions}
Based on the results obtained, the following practical recommendations can be formulated:
\begin{itemize}
\item \textbf{Use a combined dataset (data fusion)} to train models: This allows for increased stability and quality of predictions on different DEXs.
\item \textbf{Choose an approach depending on the task:}
\begin{itemize}
\item For early detection of rug pull (cessation of activity) for the purpose of buying and selling tokens, use the Idle approach.
\item For detection of fraud with a sharp drain of liquidity, use the TVL approach.
\end{itemize}
\item \textbf{Use GradientBoosting and ExtremeGradientBoosting models} as the most effective for tasks with class imbalance and high feature dimensionality.
\item \textbf{Regularly update the dataset} and train the models to adapt to changing market conditions and new fraud schemes.
\item \textbf{Implement automatic monitoring of new tokens} based on proposed models to protect investors and increase trust in the TON DeFi ecosystem.
\end{itemize}

\subsection*{Limitations}

The study identified key limitations that affect the quality and sustainability of the proposed solution. Class imbalance is one of the main problems: the number of tokens with rug pull is significantly smaller than normal tokens, which reduces the recall for the non-rug pull class and can lead to a bias in models towards the majority class. Noise and outliers in the data associated with zero values, gaps, and anomalies in transaction and price features also have a negative impact on the quality of the models, which require additional preprocessing and data filtering.

The specifics of determining rug pull depending on the chosen approach (Idle or TVL) lead to the fact that the model can miss individual cases of fraud or give false positives. Using only the first 5 minutes of trading to predict rug pull over the next hour limits the capabilities of the model on tokens with non-standard dynamics, when fraud occurs later or is of a different nature. In addition, the quality of the data received through the API or the dune.com service may be incomplete or contain errors, which also reduces the accuracy of the models and requires additional verification and validation of the information sources.

\subsection*{Future Research}

A promising direction for further research is to expand the set of features using additional metrics, such as the social activity of token creators, data from external sources, and blockchain analytics. This will increase the informativeness of the models and improve the quality of rug pull detection.

Conducting a backtest on historical data to assess the economic effect of using models in real conditions will allow a better understanding of the practical value of the proposed solution and its impact on investment returns. The use of deep learning methods that take into account time dependencies and complex patterns in data can provide an additional boost to the quality of models, especially when working with large volumes of information. Adaptation of the proposed approach to detect rug pull in other decentralized exchanges and blockchain ecosystems will expand the scope of application of research results and will help identify universal patterns of fraud in DeFi.


\bibliographystyle{elsarticle-num}
\bibliography{referencesLocal}

\end{document}